\documentclass[article,superscriptaddress,twocolumn,amsmath,amssymb,floatfix]{revtex4}
\usepackage{graphicx}% Include figure files
\usepackage{dcolumn}% Align table columns on decimal point
\usepackage{bm}% bold math
\begin{document}

\title{Interplay between topology and dynamics in the World Trade Web}
\author{D. Garlaschelli}
\affiliation{Dipartimento di Fisica, Universit\`a di Siena, Via Roma 56, 53100 Siena ITALY}
\author{T. Di Matteo}
\affiliation{Department of Applied Mathematics, Research School of Physical Sciences and Engineering, Australian National University, Canberra ACT 0200 AUSTRALIA}
\author{T. Aste}
\affiliation{Department of Applied Mathematics, Research School of Physical Sciences and Engineering, Australian National University, Canberra ACT 0200 AUSTRALIA}
\author{G. Caldarelli}
\affiliation{CNR-INFM and Dipartimento di Fisica, Universit\`a di Roma ``La Sapienza'', P.le Aldo Moro, 00183 Roma ITALY}
\author{M. I. Loffredo}
\affiliation{Dipartimento di Scienze Matematiche ed Informatiche, Universit\`a di Siena, Pian dei Mantellini 44, 53100 Siena ITALY}

%\date{\today}
\begin{abstract}
We present an empirical analysis of the network formed by the trade relationships between all world countries, or \emph{World Trade Web} (WTW). Each (directed) link is weighted by the amount of wealth flowing between two countries, and each country is characterized by the value of its \emph{Gross Domestic Product} (GDP). By analysing a set of year-by-year data covering the time interval 1950-2000, we show that the dynamics of all GDP values and the evolution of the WTW (trade flow and topology) are tightly coupled. The probability that two countries are connected depends on their GDP values, supporting recent theoretical models relating network topology to the presence of a `hidden' variable (or \emph{fitness}). On the other hand, the topology is shown to determine the GDP values due to the exchange between countries. 
This leads us to a new framework where the \emph{fitness} value is a dynamical variable determining, and at the same time depending on, network topology in a continuous feedback.
\end{abstract}
\maketitle
\section{Introduction}
The globalization process of the economy is highlighting the relevance of international interactions between world countries. The world economy is evolving towards an interconnected system of trading countries which are highly heterogeneous in terms of their internal activity. 
As many other socioeconomic and financial systems with heterogeneous interacting units, the global economy exhibits complex structure and dynamics, and can therefore be studied exploiting the tools of modern statistical mechanics.

A role of primary importance is played by the network of import/export relationships between world countries, or World Trade Web (WTW in the following). Recent empirical studies \cite{serrano,li,mywtw,myalessandria} have focused at the WTW as a complex network \cite{barabba,mendes,siam} and investigated its architecture. A range of nontrivial topological properties have been detected \cite{serrano,li,mywtw,myalessandria} and found to be tightly related to the Gross Domestic Product (GDP in the following) of world countries \cite{mywtw,myalessandria}. 
On the other hand, the economic literature has recently dealt with the study of the GDP \emph{per capita} across countries, looking for patterns and trends displayed by it. Some of these results have pointed out that the GDP \emph{per capita} displays complex dynamical behaviour and a power-law distribution across countries \cite{gallegati,gdppercapitapowerlaw}.

In the present work we extend these analyses and address the above points simultaneously. In particular, we are interested in determining empirically the effects that the WTW topology and the GDP dynamics have on each other.

\section{Data set and definitions}
The results of the present work are based on the empirical analysis of a large data set \cite{data} reporting the annual values of the population size $p_i(t)$ and of the GDP \emph{per capita} $z_i(t)$ of each world country $i$ for each year $t$ from 1948 to 2000, together with the annual amount of money $f_{ij}(t)$ flowing from each country $i$ to each country $j$ due to exports from $j$ to $i$ between 1950 and 2000. Trade, population and GDP figures are evaluated at the end of each year. 
Since in the following we are interested in a simultaneous study of GDP and trade data for each country, we restrict our analysis to the time interval 1950-2000 when both sources of information are available. We therefore set our initial time to $t_0=1950$. 
Our main interest is the total economic activity of each country, therefore we multiply each GDP \emph{per capita} $z_i(t)$ by the population size $p_i(t)$ of the corresponding country to obtain the total GDP $w_i(t)\equiv z_i(t)p_i(t)$ of that country.

\begin{figure}[]	% in second brace, h=here, t=top, b=bottom
\includegraphics[width=.48\textwidth]{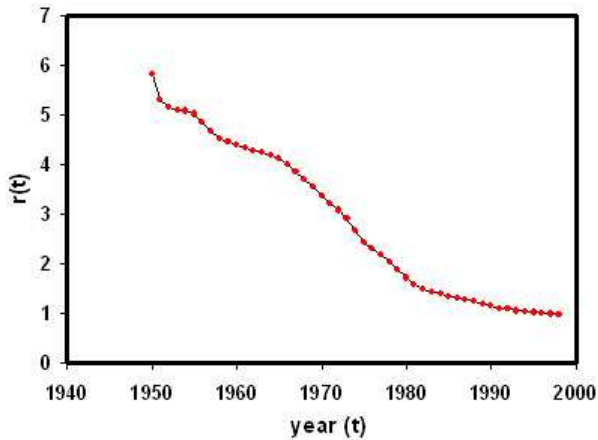}
\caption{
\label{fig_inflation}
\small Temporal evolution of the ratio $r(t)\equiv \$_t/\$_{1996}$ from $t=1950$ to $t=2000$.}
\end{figure}

Trade data are expressed in current U.S. dollars (that is, in terms of the value $\$_t$ of one U.S. dollar in the reported year $t$), while GDP data are available in current as well as in 1996 U.S. dollars ($\$_{1996}$). The use of a standard reference money unit such as $\$_{1996}$ factors out the effects of inflation and allows a comparison between figures corresponding to different years. Therefore in the following we rescale the trade data corresponding to each year $t$ to 1996 dollars. This means that $w_i(t)$ will represent the \emph{real} GDP of country $i$. A curve of the time dependence of the ratio $r(t)\equiv \$_t/\$_{1996}$ of the value of current U.S. dollars to their 1996 reference value is shown in fig. \ref{fig_inflation}. In what follows, both $f_{ij}(t)$ and $w_i(t)$ will be expressed in millions of 1996 U.S. dollars (M$\$_{1996}$).
%%%%%%%%%%%%%%%%% fig inflation %%%%%%%%%%%%%%%%%%% 
%%%%%%%%%%%%%%%%%%%%%%%%%%%%%%%%%%%%%%%%%%%%%%%%%%% 

\begin{figure}
\includegraphics[width=.48\textwidth]{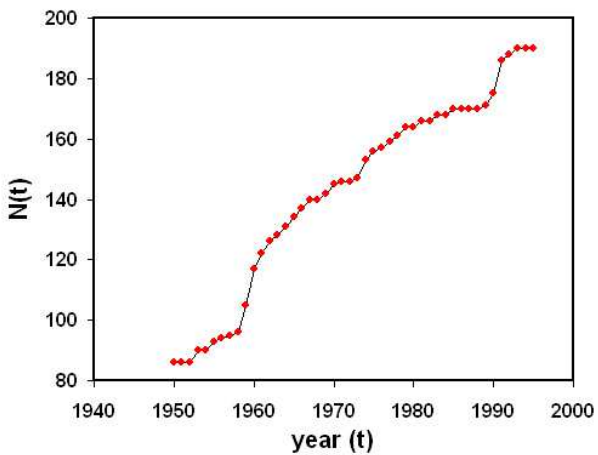}
\caption{
\label{fig_N}
\small Temporal evolution of the number of world countries $N(t)$ from $t=1950$ to $t=2000$.}
\end{figure}

The number $N(t)$ of world countries is monotonically increasing in time during the considered time interval, and it grows from $N(1950)=86$ to $N(2000)=190$. This means that the WTW is a \emph{growing network}. In fig. \ref{fig_N} we plot the time evolution of $N(t)$. The reason for the increase of $N(t)$ between 1960 and 1990 is mainly due to the declaration of independence of several countries, while the steep increase around 1990-1992 is due to the origin of many separate states from the former Soviet Union.
%%%%%%%%%%%%%%%%% fig N %%%%%%%%%%%%%%%%%%% 
%%%%%%%%%%%%%%%%%%%%%%%%%%%%%%%%%%%%%%%%% 

The total trade value of exports and imports by $i$ to/from all other countries will be denoted by $f^{in}(t)$ and $f^{out}(t)$ respectively, and it can be expressed as
\begin{eqnarray}
f^{in}_i(t)\equiv\sum_{j=1}^{N(t)}f_{ji}(t)\label{fin}\\
f^{out}_i(t)\equiv\sum_{j=1}^{N(t)}f_{ij}(t)\label{fout}
\end{eqnarray}
The net amount of incoming money due to the trading activity is therefore given by
\begin{equation}
F_i(t)\equiv f^{in}_i(t)-f^{out}_i(t)
\label{F}
\end{equation}
We finally define the adjacency matrix elements as
\begin{equation}\label{eq_adj}
a_{ij}(t)\equiv\left\{\begin{array}{lll}
1&\textrm{if}& f_{ij}(t)>0\\
0&\textrm{if}& f_{ij}(t)=0
\end{array}
\right.
\end{equation}
and the \emph{in-degree} $k^{in}_i(t)$ and \emph{out-degree} $k^{out}_i(t)$ of a country $i$ at time $t$ as
\begin{eqnarray}
k^{in}_i(t)\equiv\sum_{j=1}^{N(t)}a_{ji}(t)\\
k^{out}_i(t)\equiv\sum_{j=1}^{N(t)}a_{ij}(t)
\end{eqnarray}
representing the number of countries to which $i$ exports and from which $i$ imports respectively.
%{}`

\section{GDP: definition and empirical properties}
The Gross Domestic Product $w_i$ of a country $i$ is defined as (see for example {\tt www.investorwords.com}) the \emph{``total market value of all final goods and services produced in a country in a given period, equal to total consumer, investment and government spending, plus the value of exports, minus the value of imports"}.
In other words, there are two main terms contributing to the observed value of the GDP $w_i$ of a country $i$: an \emph{endogenous term} $I_i$ (also konwn as \emph{internal demand}) determined by the internal spending due to the country's economic process and an \emph{exogenous} term $F_i$ determined by the trade flow with other countries. The above definition can then be rephrased as
\begin{equation}\label{eq_gdpdef}
w_i(t)\equiv I_i(t)+F_i(t)
\end{equation}
where $F_i(t)$ is defined by eqs.(\ref{fin}), (\ref{fout}) and (\ref{F}). The above definition anticipates that the GDP is strongly affected by the structure of the WTW. The characterization of the interplay between the GDP dynamics and the WTW topology is the main subject of the present work. 
Before addressing this issue in detail, we first report in this section some empirical properties of the GDP.

\begin{figure}
\includegraphics[width=.48\textwidth]{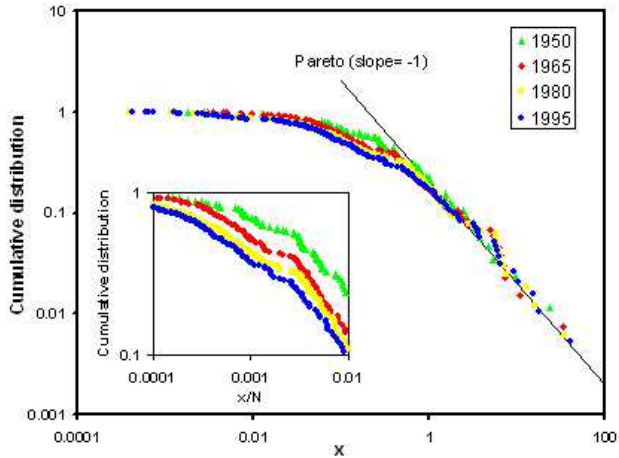}
\caption{\label{fig_gdpdistr}
\small Normalized cumulative distribution of the relative GDP $x_i(t)\equiv w_i(t)/\langle w\rangle (t)$ for all world countries at four different snapshots. Inset: the same data plotted in terms of the rescaled quantity $y_i(t)\equiv w_i(t)/w_T(t)=x_i(t)/N(t)$ for the transition region to a power-law curve.}
\end{figure}

A fundamental macroeconomic question is: how is the GDP  distributed across world countries? To address this point we consider the distribution of the rescaled quantity 
\begin{equation}\label{eq_x}
x_i(t)\equiv \frac{w_i(t)}{\langle w(t)\rangle}
\end{equation}  
where $\langle w\rangle\equiv w_T/N$ is the average GDP and $w_T(t)\equiv\sum_{i=1}^{N(t)}w_i(t)$ is the total one.
In fig. (\ref{fig_gdpdistr}) we report the \emph{cumulative} distribution 
\begin{equation}
\rho_>(x)\equiv\int_x^{\infty}\rho(x')dx'
\end{equation}
for four different years in the time interval considered. The right tail of the distribution roughly follows a straight line in log-log axes, corresponding to a power-law curve
\begin{equation}
\rho_>(x)\propto x^{1-\alpha}
\end{equation}
with exponent $1-\alpha=-1$, which indicates a tail in the GDP probability distribution $\rho(x)\propto x^{-\alpha}$ with $\alpha=2$. This behaviour is qualitatively similar to the power-law character of the \emph{per capita} GDP distribution \cite{gallegati,gdppercapitapowerlaw}.

Moreover, it can be seen that the cumulative distribution departs from the power-law behaviour in the small $x$ region, and that the value $x^*$ where this happens is larger as time increases. However, if $x_i(t)$ is rescaled to 
\begin{equation}\label{eq_y}
y_i(t)\equiv \frac{w_i(t)}{w_T(t)}=\frac{x_i(t)}{N(t)}
\end{equation}
then the point $y^*=x^*/N\approx 0.003$ where the power-law tail of the distribution starts is approximately constant in time (see inset of fig. \ref{fig_gdpdistr}). This suggests that the temporal change of $x^*$ is due to the variation of $N(t)$ affecting $\langle w(t)\rangle$ and not to other factors. 
This is because the temporal variation of $N(t)$ affects the average-dependent quantities such as $x$:
note that, while for a system with a fixed number $N$ of units $w_T$ would be simply proportional to the average value $\langle w\rangle$, for our system with a varying number of countries the two quantities can be very different. In particular, the average values of the quantities of interest may display sudden jumps due to the steep increase of $N(t)$ rather than to genuine variations of the quantities themselves.

%%%%%%%%%%%%%%%%% fig gdpdistr %%%%%%%%%%%%%%%%%%% 
%%%%%%%%%%%%%%%%%%%%%%%%%%%%%%%%%%%%%%%%% 

\section{Effects of the GDP on the WTW}
In a recent work \cite{mywtw} it was shown that the topology of the WTW, which is encapsulated in its adjacency matrix $a_{ij}$ defined in eq.(\ref{eq_adj}), strongly depends on the GDP values $w_i$. Indeed, the problem can be mapped onto the so-called \emph{fitness model} \cite{fitness,fitness2} 
where it is assumed that the probability $p_{ij}$ for a link from $i$ to $j$ is a function  $p(x_i,x_j)$ of the values of a \emph{fitness} variable $x$ assigned to each vertex and drawn from a given distribution. The importance of this model relies in the possibility to write all the expected topological properties of the network (whose specification requires in principle the knowledge of the $N^2$ entries of its adjacency matrix) in terms of only $N$ fitness values. 
Several topological properties including the degree distribution, the degree correlations and the clustering hierarchy were shown to be determined by the GDP distribution \cite{mywtw}. Moreover, an additional understanding of the WTW as a directed network comes from the study of its reciprocity \cite{myreciprocity}, which represents the strong tendency of the network to form pairs of mutual links pointing in opposite directions between two vertices. In this case too, the observed reciprocity structure can be traced back to the GDP values \cite{myreciprocity2}. 
All these results were also shown to be robust in time and to be displayed by all snapshots of the WTW \cite{myalessandria}. In this section we summarize and further extend these analyses. 

Combining the results presented in refs.\cite{mywtw,myreciprocity,myreciprocity2}, the probability that at time $t$ a link exists from $i$ to $j$ ($a_{ij}=1$) is empirically found to be 
\begin{equation}\label{fitness}
p_{t}[x_i(t),x_j(t)]=\frac{\alpha(t) x_i(t)x_j(t)}{1+\beta(t) x_i(t)x_j(t)}
\end{equation}
where $x_i$ is the rescaled GDP defined in eq.(\ref{eq_x}) and the parameters $\alpha(t)$ and $\beta(t)$ can be fixed by imposing that the expected number of links
\begin{equation}\label{expL}
L_{exp}(t)=\sum_{i\ne j}p_{t}[x_i(t),x_j(t)]
\end{equation}
equals its empirical value \cite{mywtw}
\begin{equation}
L(t)=\sum_{i\ne j}a_{ij}(t)
\end{equation}
and that the expected number of \emph{reciprocated} links \cite{myreciprocity,myreciprocity2}
\begin{equation}\label{expLrep}
L^\leftrightarrow_{exp}(t)=\sum_{i\ne j}p_{t}[x_i(t),x_j(t)]p_{t}[x_j(t),x_i(t)]
\end{equation}
equals its observed value \cite{mywtw,myalessandria,myreciprocity,myreciprocity2}
\begin{equation}
L^\leftrightarrow(t)=\sum_{i\ne j}a_{ij}(t)a_{ji}(t)
\end{equation} 

\begin{figure}
\includegraphics[width=.48\textwidth]{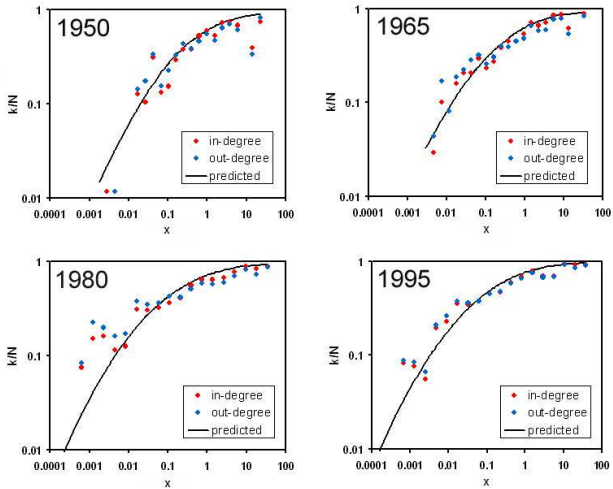}
\caption{
\label{fig_k(x)}
\small Plot of the rescaled degrees versus the rescaled GDP at four different years, and comparison with the expected trend.}
\end{figure}

\begin{figure}[ht]	% in second brace, h=here, t=top, b=bottom
\includegraphics[width=.48\textwidth]{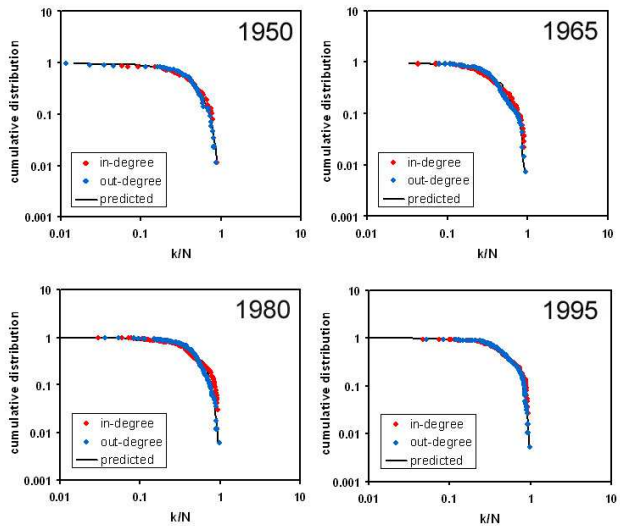}
\caption{
\label{fig_p(k)}
\small Cumulative degree distributions of the WTW for four different years and comparison with the expected trend.}
\end{figure}

This particular structure of the WTW topology can be tested by comparing various expected topological properties with the empirical ones. For instance, we can compare the empirical and the theoretical plots of vertex degrees (at time $t$) versus their rescaled GDP $x_i(t)$ \cite{mywtw}.
Note that since $p_{t}[x_i(t),x_j(t)]$ is symmetric under the exchange of $i$ and $j$, at any given time the expected in-degree and the expected out-degree of a vertex $i$ are equal (and this is indeed observed in real data, as we now show). We simply denote both by $k^{exp}_i$, which can be expressed as
\begin{equation}\label{expdegree}
k^{exp}_i(t)=\sum_{j\ne i}p_{t}[x_i(t),x_j(t)]
\end{equation}
Since the number of countries $N(t)$ increases in time, we define the rescaled degrees $\tilde{k}_i(t)\equiv k_i(t)/[N(t)-1]$ that always represent the fraction of vertices which are connected to $i$ (the term $-1$ comes from the fact that there are no self-loops in the network, hence the maximum degree is always $N-1$). In this way, we can easily compare the data corresponding to different years and network sizes.
The results are shown in figs.\ref{fig_k(x)} for various snapshots of the system. The empirical trends are in accordance with the expected ones. 
%%%%%%%%%%%%%%%%% fig k(x) %%%%%%%%%%%%%%%%%%% 
%%%%%%%%%%%%%%%%%%%%%%%%%%%%%%%%%%%%%%%%% 
Then we can also compare the cumulative distribution $P_>^{exp}(\tilde{k}^{exp})$ of the expected degrees with the empirical degree distributions $P_>^{in}(\tilde{k}^{in})$ and $P_>^{out}(\tilde{k}^{out})$. The results are shown in fig.\ref{fig_p(k)}. They confirm a good agreement between the theoretical prediction and the observed behaviour.

Note that the accordance with the predicted behaviour is extremely important since the expected quantities are computed by using only the $N$ GDP values of all countries, with no information regarding the $N^2$ trade values. On the other hand, the empirical properties of the WTW topology are extracted from trade data, with no knowledge of the GDP values. The agreement between the properties obtained by using these two independent sources of information is therefore surprising. Also note that all the sums in eqs.(\ref{expL}), (\ref{expLrep}) and (\ref{expdegree}) can be rewritten in terms of integrals involving only $p_t[x_i(t),x_j(t)]$ and $\rho(x)$ \cite{fitness}. The same is true for any other expected topological property \cite{fitness,fitness2}. This shows very clearly that the WTW topology crucially depends on the GDP distribution $\rho(x)$ shown in fig.\ref{fig_gdpdistr}.

%%%%%%%%%%%%%%%%% fig p(k) %%%%%%%%%%%%%%%%%%% 
%%%%%%%%%%%%%%%%%%%%%%%%%%%%%%%%%%%%%%%%% 

\section{Effects of the WTW on the GDP}
The above results show that the GDP and its distribution determine many topological properties of the WTW. As we anticipated, eqs.(\ref{F}) and (\ref{eq_gdpdef}) suggest that the opposite is also true, since the GDP is determined by the flow matrix $f_{ij}$ representing the WTW as a weighted network.
Understanding the detailed dependence of the GDP on the WTW is an intriguing but difficult problem that is currently unsolved.
Here we suggest a first step towards its solution by proposing a possible framework to be further explored in the future. 

The way the structure of the WTW affects the GDP through eq.(\ref{eq_gdpdef}) is twofold: firstly, its topology determines which are the nonzero entries of the adjacency matrix $a_{ij}(t)$ and therefore of $f_{ij}(t)$; secondly, the nonzero weights $f_{ij}(t)$ themselves determine $w_i(t)$ quantitatively through eq.(\ref{F}). From the study of dynamical processes on complex networks it is now understood in general that the former, purely topological aspect has a primary qualitative impact on the collective dynamical behaviour, while the latter has mainly a quantitative effect on the dynamics. 
Now we show an example of the crucial role of the topology in a paradigmatic case which also turns out to be a good candidate for the modeling of our system. 

Interestingly, the effects of a nontrivial network topology on the dynamics of wealth exchange has been addressed theoretically in a series of papers \cite{BM,souma,tizianawealth,mypavia}. In these works, the exchange of wealth is modeled through a stochastic process taking place on an underlying network of agents. The most general choice which has been proposed \cite{tizianawealth} for such a process is the following evolution rule for the wealth $w_i$ of the $i$-th agent (with $i=1,N$):
\begin{eqnarray}
w_i(t+1)&=&w_i(t)+\xi_i(t)+\eta_i(t)w_i(t)+\nonumber \\ 
&+&\sum_j[J_{ji}(t)w_j(t)-J_{ij}(t)w_i(t)]
\label{eq_model}
\end{eqnarray}
where $w_i(t)$ denotes the wealth of agent $i$ at time $t$, $\xi_i(t)$ is an additive noise term, $\eta_i(t)$ is a multiplicative noise term, and $J_{ij}(t)$ represents the fraction of agent $i$'s wealth being transferred to agent $j$. 
The above model is therefore the combination of an additive and of a multiplicative stochastic process occurring at discrete time steps among the $N$ agents. 
It is general enough to suspect that the combined dynamics of the GDP and the WTW can be captured successfully by it. We now discuss how the results presented in the literature can be exploited to gain insight into this problem. 

For the above model to be fully defined, one needs to specify the probability distribution for the stochastic variables $\xi_i(t)$ and $\eta_i(t)$ as well as the matrix elements $J_{ij}(t)$. Several choices have been explored in the literature. The most studied case is the purely multiplicative one where $\xi_i(t)\equiv0$ and $\eta_i(t)$ is a Gaussian noise term \cite{BM,souma,mypavia}. The opposite, purely additive case where $\eta_i(t)\equiv 0$ and $\xi_i(t)$ is a Gaussian variable has also been considered \cite{tizianawealth}. The choices for $J_{ij}(t)$ in all cases assume the following dependence on the adjacency matrix $a_{ij}(t)$ of the underlying network:
\begin{equation}
J_{ij}(t)=q_{ij}(t)a_{ij}(t)=\left\{
\begin{array}{lll}
q_{ij}(t)&\textrm{if}& a_{ij}(t)=1\\
0&\textrm{if}&a_{ij}(t)=0
\end{array}
\right.
\end{equation}
The role of $q_{ij}(t)$ is to distribute the wealth coming out from $i$ among its neighbours. If all agents deliver the same fraction of their wealth to each of their neighbours, the choice $q_{ij}=q/N$ with constant $q$ is used \cite{BM,mypavia}. The choices $q_{ij}=q/k^{in}_j$ \cite{souma} and $q_{ij}=q/k^{out}_i$ \cite{tizianawealth} are instead considered if the total wealth  respectively received or delivered by each agent is a fixed fraction $q$ of its wealth, which is equally distributed among its neighbours. 
The main point is then the specification of $a_{ij}(t)$, on which also $q_{ij}(t)$ depends. Here the crucial role of the topology emerges. Several choices has been explored, but in all cases assuming that the topology is held fixed: $a_{ij}(t)=a_{ij}$. 
In the purely multiplicative case $\xi_i(t)\equiv 0$ it was shown \cite{BM} that on fully connected graphs ($a_{ij}=1$ $\forall i,j$) the rescaled wealth $x_i\equiv w_i/\langle w\rangle$ eventually approaches a stationary distribution displaying a power-law tail with exponent determined by $q$ and by the variance of the distribution of the multiplicative term $\eta_i(t)$. Note that in this case all the above choices for $q_{ij}$ become equivalent since $k_i=N-1\approx N$. This also implies that $J_{ij}(t)=q/N$ $\forall i,j$. The fully connected case is therefore a prototypic example showing that, irrespective of the details, the wealth distribution approaches a power-law curve.
The opposite possibility is that of an empty graph where all vertices are disconnected from each other: $a_{ij}=0$ $\forall i,j$. This clearly implies that $J_{ij}(t)=0$ $\forall i,j$. In the purely multiplicative case, this model is easily shown to generate a log-normal wealth distribution since $\log w_i$ is a sum of independent identically distributed random variables, eventually approaching a Gaussian distribution as ensured by the Central Limit Theorem. Therefore this is the opposite paradigmatic case where, irrespective of the model details, the wealth distribution displays a log-normal form.

Interestingly, many empirical wealth and income distributions display a mixed shape with a power-law tail in the large wealth region and a different behaviour (which in some cases is log-normal-like) for the small wealth range  \cite{tizianawealth,tizefabio,tizeyako}. 
This kind of behaviour is also displayed by the GDP distribution across world countries as shown in fig.\ref{fig_gdpdistr}.
Since all real networks fall somewhere in between fully connected and empty graphs, it is interesting to ask whether the observed mixed shape of wealth distributions can be accounted for by the topological properties of the underlying network.
This has stimulated the exploration of the model defined in eq.(\ref{eq_model}) in the case of a nontrivial topology \cite{souma,tizianawealth,mypavia}. Remarkably, in the case of purely additive noise on scale-free networks \cite{tizianawealth} and of purely multiplicative noise on heterogeneous networks with varying link density \cite{mypavia}, the wealth distribution has been shown to approach a shape very similar to the observed one. 

The above results allow to figure out a likely mechanism for the GDP evolution driven by the WTW structure. 
Combining eqs.(\ref{fin},\ref{fout},\ref{F},\ref{eq_gdpdef}) yields the explicit definition of GDP in terms of the import/export terms:
\begin{equation}
w_i(t)=I_i(t)+\sum_{j}[f_{ji}(t)-f_{ij}(t)]
\end{equation}
It is instructive to compare the above expression with the model defined in eq.(\ref{eq_model}). 
First of all, note that the GDP is evaluated and publicly released quarterly as the main indicator of the overall economic activity of a country. Based on the trend exhibited by the GDP at the end of each period, each country plans suitable measures to adjust its  economic activity during the following period. This means that the GDP evolves through discrete timesteps in a way similar to the update rule defined in eq.(\ref{eq_model}). 
Then, note that the internal demand $I_i(t)$ is an endogenous term that does not depend on the GDP of other countries, while instead the trade values $f_{ij}(t)$ may depend on both $w_i(t)$ and $w_j(t)$. Therefore the evolution of $w_i(t)$ must be the combination of an endogenous and an exogenous dynamics. 
Finally, it is reasonable to suspect that, due to the multiplicative character of the economy (each country reinvests the income of the previous period in the internal activity as well as in the trade relationships), the dependece of both the internal demand $I_i(t)$ and the trade matrix $f_{ij}(t)$ on the GDP values is linear as in eq.(\ref{eq_model}) with $\eta_i(t)\ne 0$. Therefore we conjecture that a model similar to eq.(\ref{eq_model}) may capture the basic properties of the dynamics of the GDP driven by the WTW structure. This possibility will be explored in detail in future papers. 

Of course, in a realistic model for the coupled GDP and WTW evolution, the strong assumption of a fixed topology ($a_{ij}(t)=a_{ij}$) that has been so far used in the literature must be relaxed, and the time-dependence of the interaction matrix $J_{ij}(t)$ in eq.(\ref{eq_model}) fully exploited. 
On the other hand, we showed that the topology of the WTW is at each timestep well predicted by the knowledge of the values $\{w_i\}_i$ as evident from eqs.(\ref{fitness},\ref{expL},\ref{expLrep},\ref{expdegree}). Therefore the evolution of $a_{ij}(t)$ is not independent from the GDP values, and it should instead contain an explicit dependence on them. This dependence should then be plugged into the term $J_{ij}(t)$ in eq.(\ref{eq_model}). Viewed from the point of view of network theory, this property leads us to a novel framework where the network is shaped at each timestep by the set of \emph{fitness} values \cite{fitness}, which in turn evolve in a topology-dependent way through a stochastic dynamics similar to that in eq.(\ref{eq_model}). The extension of the current fitness formalism to take these evolutionary aspects into account is an intriguing problem to address in the future. 

\section{Conclusions}
In the present work we have reported a range of empirical results and theoretical arguments highlighting the interplay between the dynamics of the GDP and the topology of the WTW. The topological properties of the trade network have been shown to be determined by the GDP of all world countries. On the other hand, the 
empirical properties of the GDP distribution across world countries can be traced back to the underlying dynamical process of wealth exchange among countries, which is mediated by the WTW. 
The emerging picture is that of a discrete process where at each timestep the GDP distribution is determined by the WTW as a weighted network, and at the same time the WTW topology is also determined by the GDP values. 
We have thus suggested the need for a theoretical framework where the network topology is determined by some hidden quantity, which is in turn an evolving variable (and not simply a quenched one) whose dynamics is coupled to the topology. The present work provides robust empirical evidence for such a framework, highlighting the need to further develop current models in order to take evolutionary aspects into account. We have proposed a paradigmatic class of stochastic models for the fitness variable that may suggest a possible way to integrate the dynamics with the topology. The present work represents a basis for a future understanding of the details of this interplay in the case of the world economy. Such an improved framework would give predictive results on extremely important issues such as the economic power and interdependence of world countries, but also on many other problems in social science and biology where formally similar mechanisms are at work.\\

{\bf Acknowledgments }\\
TDM and TA wish to thank the partial support by ARC Discovery Projects:
DP03440044 (2003), DP0558183 (2005) and COST P10 ``Physics of Risk'' project and M.I.U.R.-F.I.S.R. Project ``Ultra-high frequency dynamics of financial markets''.

\end{document}